 \documentclass[preprint2]{aastex} 

\slugcomment{}

\shorttitle{The Solar Cycle: A new prediction technique based on
logarithmic values}
  \shortauthors{Z. L. Du}

\begin{document}
\title{The Solar Cycle: A new prediction technique based on
logarithmic values}

\author{Z. L. Du\altaffilmark{}}
\affil{Key Laboratory of Solar Activity, National Astronomical
Observatories, Chinese Academy of Sciences, Beijing 100012, China}
\email{zldu@nao.cas.cn}

\begin{abstract}
A new prediction technique based on logarithmic values is proposed
to predict the maximum amplitude ($R_{\mathrm{m}}$) of a solar
cycle from the preceding minimum $aa$ geomagnetic index
($aa_{\mathrm{min}}$). The correlation between
$\mathbf{ln}R_{\mathrm{m}}$ and $\mathbf{ln}aa_{\mathrm{min}}$
($r=0.92$) is slightly stronger than that between $R_{\mathrm{m}}$
and $aa_{\mathrm{min}}$ ($r=0.90$). From this method, cycle 24 is
predicted to have a peak size of
$R_{\mathrm{m}}(24)=81.7(1\pm13.2\%)$. If the suggested error in
$aa$ (3 nT) before 1957 is corrected, the correlation coefficient
between $R_{\mathrm{m}}$ and $aa_{\mathrm{min}}$ ($r=0.94$) will
be slightly higher, and the peak of cycle 24 is predicted much
lower, $R_{\mathrm{m}}(24)=52.5\pm13.1$. Therefore, the prediction
of $R_{\mathrm{m}}$ based on the relationship between
$R_{\mathrm{m}}$ and $aa_{\mathrm{min}}$ depends greatly on the
accurate measurement of $aa$.
\end{abstract}


\section{Introduction}           
\label{sect:intro}

Predicting the strength of an upcoming solar cycle
($R_{\mathrm{m}}$) is important in both solar physics and space
weather. A variety of methods have been used to do so, of which
some are based on statistics and some others are related to
physics
\citep{Kane07a,Cameron07,Pesnell08,Hiremath08,Tlatov09,Messerotti09,Petrovay10,Du10}.
A reliable prediction of $R_{\mathrm{m}}$ may test models for
explaining the solar cycle \citep{Pesnell08}. Various solar dynamo
models \citep[e.g.,][]{Dikpati06,Choudhuri07} have been proposed
to explain the solar cycle but the predictive skill of
$R_{\mathrm{m}}$ needs to be checked in future
\citep{Cameron07,Pesnell08,Du11a}. Based on the Solar Dynamo
Amplitude (SODA) index, \citet{Schatten05} predicted that the peak
sunspot number of the current cycle (24) will be low, at $\sim80$.
\citet{Dikpati06} predicted that the peak size of cycle 24 will be
30\,\%--\,50\% higher than that of cycle 23 based on a modified
flux-transport dynamo model. In contrast, \citet{Choudhuri07}
predicted that the peak size of cycle 24 will be 30\,\%--\,50\%
lower than that of cycle 23 based on a flux-transport dynamo
model.

Since \citet{Ohl66} found a high correlation between the minimum
$aa$ geomagnetic activity ($aa_{\mathrm{min}}$) in the declining
phase of a solar cycle and the maximum sunspot number of the
succeeding cycle ($R_{\mathrm{m}}$), a great many papers related
to this finding have been published over the past decades
\citep{Brown69,Kane10,Wilson90,Hathaway06,Charvatova09,Wang09}.
The level of geomagnetic activity near the time of solar activity
minimum has been shown to be a good indicator for the amplitude of
the following solar activity maximum
\citep{Ohl76,Wilson90,Layden91,Thompson93,Hathaway06,Kane10}. This
method based on a solar dynamo concept that the geomagnetic
activity during the declining phase of the preceding cycle or at
the cycle minimum provides approximately a measure of the poloidal
solar magnetic field that generates the toroidal field for the
next cycle \citep{Schatten78}.

When geomagnetic precursor methods are applied to the current
cycle (24), some discrepancies are shown for different authors.
\citet{Hathaway06} predicted $R_{\rm m}(24)=160\pm25$ using the I
component of $aa$ by subtracting the linear R component with
$R_{\rm m}$ from $aa$ \citep{Feynman82}. \citet{Dabas08} employed
the number of geomagnetic disturbed days 
prior to the minimum of the sunspot cycle, and predict $R_{\rm
m}(24)=124\pm23$. \citet{Wang09} predicted $R_{\rm m}(24)=97\pm25$
based on the total open flux at sunspot minimum, which is derived
from the historical $aa$ index by removing the contribution of the
solar wind speed.

The correlation coefficients between $R_{\mathrm{m}}$ and the
geomagnetic-based parameters are usually very high, from 0.8
\citep{Ohl76,Wilson90,Thompson93,Shastri98,Kane10} up to 0.97
\citep{Layden91,Lantos98,Hathaway99,Dabas08}. However, a high
correlation coefficient does not always yield an accurate
prediction, such as in the case of cycle 23 
\citep{Kane07a}. It was found that the correlation coefficient
between $R_{\mathrm{m}}$ and $aa_{\mathrm{min}}$ varies roughly in
a cycle of about 44-year and that the prediction error based on
this method when the correlation coefficient decreases is much
larger than that when the correlation coefficient increases
\citep{Du309b,Du11a}.

Conventionally, the correlation between $R_{\mathrm{m}}$ and
$aa_{\mathrm{min}}$ is analyzed by a linear relationship (Section
\ref{sec:Linear}), and cycle 19 is viewed as anomalous or an
`outlier' due to the very great $R_{\mathrm{m}}$ ever seen.
However, by analyzing the relationship between the logarithms of
$R_{\mathrm{m}}$ and $aa_{\mathrm{min}}$ in Section
\ref{sec:logCorrelation}, cycle 19 is no longer anomalous from the
scatter points of $\mathbf{ln}R_{\mathrm{m}}$ versus
$\mathbf{ln}aa_{\mathrm{min}}$. Whether correcting the suggested
error (3 nT) in aa before 1957 has great influences on the
prediction of $R_{\mathrm{m}}$ based on the relationship between
$R_{\mathrm{m}}$ and $aa_{\mathrm{min}}$ (Section
\ref{sec:Correction}). The results are briefly
discussed and summarized in Section \ref{sec:Discussions}. 

\section{Linear relationship between $R_{\mathrm{m}}$ and
$aa_{\mathrm{min}}$} \label{sec:Linear}

This study uses the annual values of geomagnetic $aa$ index
computed from the 3-hourly $K$ indices at two near-antipodal
midlatitude stations~\citep{Mayaud72,Love11} since
1868\footnote{ftp://ftp.ngdc.noaa.gov/STP/SOLAR\_DATA/RELATED\-\_INDICES/AA\_INDEX/}
and the equivalent ones from measurements taken in Finland   
from 1844 to 1867~\citep{Nevanlinna93,Nevanlinna04}, and the
annual values of the International sunspot number ($R_{\rm z}$)
since 1844\footnote{http://www.sidc.be/sunspot-data/} produced by
the Solar Influences Data Analysis Center (SIDC), World Data
Center for the Sunspot Index, at the Royal Observatory of Belgium.
The maximum amplitude of sunspot cycle ($R_{\mathrm{m}}$) and the
preceding $aa$ minimum ($aa_{\mathrm{min}}$) are listed in
Table~\ref{Tab:1}.

\begin{figure}[!tb]
 \includegraphics[width=\columnwidth]{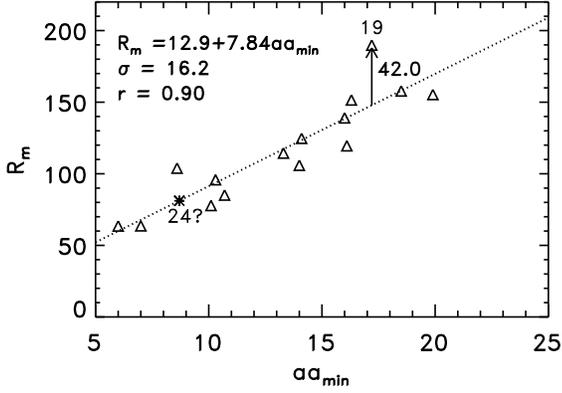}
 \caption{ Scatter plot of $R_{\mathrm{m}}$ {\it vs.} $aa_{\rm min}$
     (triangles)
     } 
 \label{Fig:1}
 \end{figure}

Conveniently, one employs the linear relationship between
$R_{\mathrm{m}}$ and $aa_{\mathrm{min}}$ to predict the former.
Figure~\ref{Fig:1} depicts the scatter plot between
$R_{\mathrm{m}}$ and $aa_{\rm min}$ (triangles). The dotted line
indicates the linear fit of $R_{\mathrm{m}}$ to
$aa_{\mathrm{min}}$,
\begin{equation}
  \label{Eq:linear}
  R_{\mathrm{m}}=12.9\pm14.7+(7.84\pm1.06)aa_{\mathrm{min}},
\end{equation}
where the values following $\pm$ indicate the standard deviation.
The correlation coefficient between $R_{\mathrm{m}}$ and
$aa_{\mathrm{min}}$ is very high, $r=0.90$ at the 99\% level of
confidence. From this equation and $aa_{\mathrm{min}}(24)=8.7$,
the peak size of cycle 24 is predicted to be
$R_{\mathrm{m}}(24)=81.2\pm 16.2$, where $\sigma=16.2$ is the
standard deviation of fitting, defined by
\begin{equation}
  \label{Eq: }
  \sigma=\sqrt{\frac{\sum_{i=9}^{23}[R_{\mathrm{f}}(i)-R_{\mathrm{m}}(i)]^2}{N-1}},
\end{equation}
where $R_{\mathrm{f}}$ is the fitted value of $R_{\mathrm{m}}$ by
Equation~(\ref{Eq:linear}) and $N=15$ is the number of data pairs.

It is seen in Fig.~\ref{Fig:1} that the point of cycle 19 is far
above the fitting line, $\Delta R_{\mathrm{m}}(19)=42.0$.
Therefore, cycle 19 is often called as an `outlier'
\citep{Kane07a} and this point may, occasionally, be deleted in
order to obtain a better correlation.

\section{Relationship between $\mathbf{ln}R_{\mathrm{m}}$ and
$\mathbf{ln}aa_{\mathrm{min}}$} \label{sec:logCorrelation}

\begin{figure}[!tb]
 \includegraphics[width=\columnwidth]{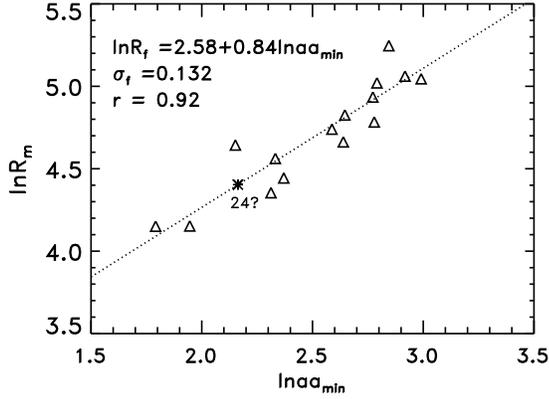}
 \caption{Scatter plot of $\mathbf{ln} R_{\mathrm{m}}$ {\it vs.} $\mathbf{ln}aa_{\rm min}$
     (triangles).
     } 
 \label{Fig:2}
 \end{figure}

Now, we analyze the scatter plot between the 
logarithms of $R_{\mathrm{m}}$ and $aa_{\mathrm{min}}$, as shown
in Fig.~\ref{Fig:2} (triangles).

The correlation coefficient of $\mathbf{ln}R_{\mathrm{m}}$ with
$\mathbf{ln}aa_{\mathrm{min}}$ is $r=0.92$ at the 99\% level of
confidence, slightly stronger than that of $R_{\mathrm{m}}$ with
$aa_{\mathrm{min}}$ (0.90). The least-squares-fit regression
equation (dotted line) is
\begin{equation}
  \label{Eq:log}
  \mathbf{ln}R_{\mathrm{f}}=2.58\pm0.26+(0.84\pm0.10) \mathbf{ln}aa_{\mathrm{min}},
\end{equation}
where $\mathbf{ln}R_{\mathrm{f}}$ denotes the fitted value of
$\mathbf{ln}R_{\mathrm{m}}$. The standard deviation of fitting is
$\sigma_{\mathrm{f}} =0.132$. This equation is equivalent to the
form of a power-law,
\begin{equation}
  \label{Eq:power}
  R_{\mathrm{f}}=e^{2.58}aa_{\mathrm{min}}^{0.84},
\end{equation}
implying that $R_{\mathrm{m}}$ does not depend completely linearly
on $aa_{\mathrm{min}}$. The error of $\mathbf{ln}R_{\mathrm{f}}$
is
\begin{equation}
  \label{Eq:logerror}
  \varepsilon=\Delta
\mathbf{ln}R_{\mathrm{f}}=\mathbf{ln}R_{\mathrm{f}}-\mathbf{ln}R_{\mathrm{m}}.
\end{equation}
The values of $R_{\mathrm{f}}$ and $\varepsilon$ are listed in
Table~\ref{Tab:1}.

%
 \begin{table}[!tb]
 \small
 \caption{Annual 
$aa_{\mathrm{min}}$, $R_{\mathrm{m}}$ and Fitted Results} 
  \label{Tab:1}
 \begin{tabular}{rrr|rr|rr}
 \tableline  
    & \multicolumn{2}{c}{Parameters} & \multicolumn{2}{c}{From $\mathbf{ln}aa_{\mathrm{min}}$\tablenotemark{a}}
    & \multicolumn{2}{c}{Corrected $aa$\tablenotemark{b}}\\
 $n$ & $aa_{\mathrm{min}}$  & $R_{\mathrm{m}}$ & $R_{\mathrm{f}}$ &
  $|\varepsilon|(\%)$ &  $R_{\mathrm{p}}$ & $|\Delta R_{\mathrm{p}}|$\\
  \tableline
9   & 14.1 & 124.7  & 122.8 & $1.5$  & 132.7&  8.0 \\
10  & 10.3 &  95.8  &  94.2 & $1.6$  &  96.4&  0.6\\
11  & 16.0 & 139.0  & 136.6 & $1.7$  & 150.8&  11.8\\
12  &  7.0 &  63.6  &  68.0 &   6.7  &  64.9&  1.3\\
13  & 10.7 &  85.1  &  97.3 &  13.4  & 100.2& 15.1\\
14  &  6.0 &  63.5  &  59.7 & $6.1$  &  55.4& $8.1$\\
15  &  8.6 & 103.9  &  80.9 &$25.0$  &  80.2& $23.7$\\
16  & 10.1 &  77.8  &  92.7 &  17.5  &  94.5& 16.7\\
17  & 13.3 & 114.4  & 116.9 &   2.1  & 125.0& 10.6\\
18  & 16.3 & 151.5  & 138.8 & $8.8$  & 153.7& 2.2\\
19  & 17.2 & 189.8  & 145.2 &$26.8$  & 162.3& $27.5$\\
20  & 14.0 & 105.9  & 122.1 &  14.2  & 103.1& $2.8$\\
21  & 19.9 & 155.3  & 164.3 &   5.6  & 159.4& 4.1\\
22  & 18.5 & 157.8  & 154.5 & $2.1$  & 146.0& $11.8$\\
23  & 16.1 & 119.5  & 137.4 &  13.9  & 123.1& 3.6\\
\tableline
$\overline{|x|}$  &  13.2 & 116.5    & 115.4 & 9.8  & 116.5& 9.9\\
\tableline
24  &  8.7 & ?      & ?81.7 & ?13.2  & ?52.5 & ?13.1\\
 \tableline 
 \end{tabular}
%
   \tablenotetext{a}{From Fig.~\ref{Fig:2} and Eqs.~(\ref{Eq:log}) and (\ref{Eq:logerror}).}
   \tablenotetext{b}{From Fig.~\ref{Fig:3} and Eqs.~(\ref{Eq:corrected}) and (\ref{Eq:correctederror}).}
 \end{table}

One can see from Table \ref{Tab:1} that the maximum relative error
occurs in cycle 19, $|\varepsilon(19)|=26.8\%$, which is only
slightly larger than that in cycle 15, $|\varepsilon(15)|=25.0\%$.
Therefore, cycle 19 seems to be not an `outlier' as cycle 15 in
view of the relative error. In fact, \citet{Ramesh11} proved,
through a thorough analysis of the linear relationship between
$R_{\mathrm{m}}$ and the preceding sunspot minimum
($R_{\mathrm{min}}$), that cycle 19 is not an outlier --- it is
more appropriate to be called as an anomalous.

From Equation~(\ref{Eq:log}), the peak sunspot number for cycle 24
can be predicted: $\mathbf{ln}R_{\mathrm{f}}(24)=4.403\pm0.132$ or
$R_{\mathrm{f}}(24)=81.7(1\pm13.2\%)$, close to that by the linear
relationship (81.2).

\section{Using the corrected $aa$} \label{sec:Correction}

\begin{figure}[h]
\epsscale{1.0}
   \plotone{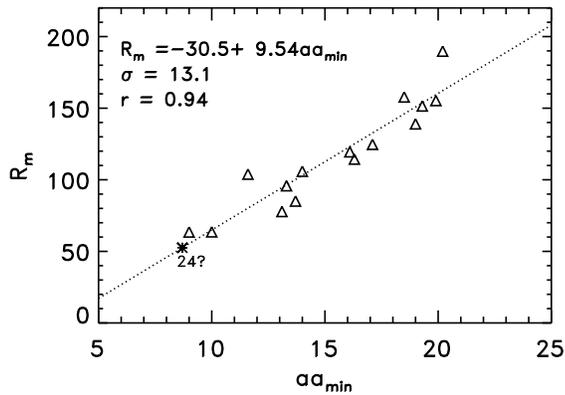}
 \figcaption{  Scatter plot of  $R_{\rm m}$ {\it vs.} the corrected $aa_{\rm min}$ (triangles)
 by adding 3 nT to $aa$ before 1957
 \label{Fig:3}}
\end{figure}

The $aa$ index was suggested to exist an error and should be
increased by 3 nT before
1957~\citep{Nevanlinna93,Lukianova09,Svalgaard04}. In this
section, we corrected the suggested error in $aa$ by adding 3 nT
to $aa_{\rm min}$ for cycles 9\,--\,19 and re-examine the previous
results. The Scatter plot between  $R_{\rm m}$ and the corrected
$aa_{\rm min}$ is shown in Fig.~\ref{Fig:3} (triangles).

The correlation coefficient between $R_{\mathrm{m}}$ and
$aa_{\mathrm{min}}$ is now $r=0.94$ at the 99\% level of
confidence, slightly stronger than that using the un-corrected
$aa_{\mathrm{min}}$ (0.90). The linear regression equation (
dotted line) is
\begin{equation}
  \label{Eq:corrected}
  R_{\mathrm{m}}=-30.5\pm15.8+(9.54\pm1.00) aa_{\mathrm{min}}, 
\end{equation}
and the standard deviation of fitting is $\sigma=13.1$.
Table~\ref{Tab:1} lists the fitted result ($R_{\mathrm{p}}$) and
the prediction error of $R_{\mathrm{p}}$,
\begin{equation}
  \label{Eq:correctederror}
  \Delta R_{\mathrm{p}}=R_{\mathrm{p}}-R_{\mathrm{m}}.
\end{equation}

From Equation~(\ref{Eq:corrected}), the peak sunspot number for
the next cycle (24) is predicted to be
$R_{\mathrm{m}}(24)=52.5\pm13.1$,
much lower than that using the un-corrected $aa_{\mathrm{min}}$ in
Section~\ref{sec:Linear} ($81.2\pm 16.2$). Therefore, the
prediction of $R_{\mathrm{m}}$ based on the relationship between
$R_{\mathrm{m}}$ and $aa_{\mathrm{min}}$ depends greatly on the
accurate measurement of $aa$, that is, whether correcting the
suggested error in $aa$ before 1957.

\section{Discussions and Conclusions} \label{sec:Discussions}

Studying the variations in the 11-yr solar cycle may help to
understand the formation and dynamo mechanism of the cycle
\citep{Parker55,Babcock61}. It has long been noted that the 11-yr
Schwabe cycle is close to the synodic period of the co-alignments
of the Earth, Venus, and Jupiter \citep{Wood75,Grandpierre96}.
However, it is uncertainty whether the planetary tidal force can
trigger the dynamo mechanism \citep{Grandpierre96} as the
acceleration due to planetary tidal force is much smaller than the
observed acceleration at the level of tachocline \citep{Jager05}. %
In the dynamo mechanism, the differential rotation in the solar
convective envelope transforms the poloidal magnetic field
structure into toroidal magnetic field structure which leads to
the formation of sunspots due to Coriolis force
\citep{Parker55,Babcock61,Schatten78,Dikpati06,Choudhuri07}.
Dynamo models can reproduce certain features of the cycle (e.g.,
sunspot butterfly diagrams), but the predictive skill of
$R_{\mathrm{m}}$ has not been checked so far
\citep{Cameron07,Pesnell08,Du11a}. As the actual observational
time series of poloidal field (available only since the mid-1970s)
is not long \citep{Choudhuri07}, the geomagnetic activity around
the cycle minimum is used as a measure to estimate the poloidal
solar magnetic field \citep{Schatten78}. \citet{Javaraiah08} found
that $R_{\mathrm{m}}$ is well correlated with the sum of the
sunspot group areas in the $0^{\mathrm{o}}$ -- $10^{\mathrm{o}}$
latitude interval both of the Sun's northern hemisphere near the
minimum of the previous cycle ($r=0.95$) and of the southern
hemisphere just after the time of the maximum of the previous
cycle cycle ($r=0.97$). Recently, \cite{Tlatov09} suggested that
the parameter $G=\Sigma(1/N_{\mathrm{g}})^2$, defined by the
number of sunspot groups $N_{\mathrm{g}}\ge1$, may
be useful for calibration 
of the residual magnetic poloidal fields, as the amplitude of $G$
is highly correlated with $R_{\mathrm{m}}$ at one and a half solar
cycles later ($r=0.96$).

Conventionally, the relationship between $R_{\mathrm{m}}$ and
$aa_{\mathrm{min}}$ is analyzed linearly. The upcoming
$R_{\mathrm{m}}$ is predicted by extrapolating the linear
regression equation from a least-squares-fit algorithm
($R_{\mathrm{f}}$), and its uncertainty is estimated by the
standard deviation ($\sigma$) as the actual prediction error
($\Delta R_{\mathrm{f}}=R_{\mathrm{f}}-R_{\mathrm{m}}$) has not
been known until the cycle is over. Thus, the prediction is
usually expressed in the form of
$R_{\mathrm{m}}=R_{\mathrm{p}}\pm\sigma$ ($2\sigma$), regarding
the 68\% (95\%) level of confidence.

Giving the uncertainty in an absolute measure ($\sigma$) is enough
in most circumstances. If a prediction error ($\Delta
R_{\mathrm{m}}$) is less than 20, this prediction is usually
thought as a successful one. In some cases, however, it may
alternately be better to describe the uncertainty in a relative
form. Suppose that the prediction errors are the same for two
predictions ($R_{\mathrm{m1}}$, $R_{\mathrm{m2}}$), $\Delta
R_{\mathrm{m}}=20$. If $R_{\mathrm{m1}}>100$, this prediction is
rather successful as its relative prediction error is less than
$20\%$. However, if $R_{\mathrm{m2}}<50$, this prediction is
terrible as its relative prediction error is larger than $40\%$.
Therefore, showing the prediction error in a relative form is a
better choice, particularly in comparison with two or more
predictions.

This study examined the relationship between
$\mathbf{ln}R_{\mathrm{m}}$ and $\mathbf{ln}aa_{\mathrm{min}}$,
with a correlation coefficient ($r=0.92$) slightly higher than
that for the linear relationship between $R_{\mathrm{m}}$ and
$aa_{\mathrm{min}}$ ($r=0.90$). The standard deviation of fitting
so obtained ($\sigma_{\mathrm{f}}$) refers directly to the
relative standard deviation. From this method, the peak sunspot
number for cycle 24 is predicted to be
$R_{\mathrm{m}}(24)=81.7(1\pm13.2\%)$, near to that from a
modified Gaussian function \citep[72$\pm11$,][]{Du11d}, that from
the sunspot minimum \citep[85$\pm17$,][]{Ramesh11}, and that from
the sum of the sunspot group areas in the $0^{\mathrm{o}}$ --
$10^{\mathrm{o}}$ latitude interval of the previous cycle
\citep[$87\pm7$,][]{Javaraiah08}. In fact, the logarithmic
$R_{\mathrm{z}}$ was often used in the studies of
\citet{Waldmeier39}.

If the suggested error in $aa$ (3 nT) before 1957
\citep{Nevanlinna93,Lukianova09,Svalgaard04} is corrected, the
correlation coefficient between $R_{\mathrm{m}}$ and
$aa_{\mathrm{min}}$ ($r=0.94$) will be slightly higher than that
using the un-corrected $aa_{\mathrm{min}}$ ($r=0.90$). From this
method, the peak sunspot number for the next cycle (24) is
predicted to be $R_{\mathrm{m}}(24)=52.5\pm13.1$. It is close to
that from long-term trends of sunspot activity
\citep[55.5,][]{Tan11}, lower than both that by an autoregressive
model \citep[$110\pm11$,][]{Hiremath08} and that by the $G$
parameter \citep[$135\pm12$,][]{Tlatov09}. The prediction (52.5)
is lower than that using the total open flux derived from the $aa$
index \citep[$97\pm25$,][]{Wang09}, that using the number of
geomagnetic disturbed days \citep[$124\pm23$,][]{Dabas08}, and
that using the I component of $aa$
\citep[$160\pm25$,][]{Hathaway06}. It should be pointed out that
the prediction using the corrected $aa$ (52.5), which is close to
that by \citet[$58.0\pm25.0$,][]{Kane10}, is much lower than that
using the uncorrected $aa$ ($81.2\pm 16.2$). Therefore, the
accurate measurement of $aa$ is crucial to predict
$R_{\mathrm{m}}$ when using the relationship between
$R_{\mathrm{m}}$ and the preceding $aa_{\mathrm{min}}$. Whether
correcting the suggested error in $aa$ before 1957 may lead to
great discrepancies in the prediction of $R_{\mathrm{m}}$ by using
the above relationship.

Accurately predicting the peak size of a upcoming sunspot cycle is
a difficulty task as the monthly sunspot numbers may have
systematic uncertainties of about 25\%
\citep{Vitinskij86}. %
In fact, the relationship between $aa$ and $R_{\rm z}$ is very
complex and the current $aa$ value may be related to the past
solar activities \citep{Du11b,Du11c}, reflecting long-term
evolution characteristics of the Sun's magnetic field
\citep{Lockwood99,Tlatov09}.

Main conclusions can be drawn as follows.
\begin{enumerate}
  \item The correlation
between $\mathbf{ln}R_{\mathrm{m}}$ and
$\mathbf{ln}aa_{\mathrm{min}}$ ($r=0.92$) is slightly stronger
than that between $R_{\mathrm{m}}$ and
$aa_{\mathrm{min}}$($r=0.90$). From this method, the
$R_{\mathrm{m}}$ for cycle 24 is predicted to be
$R_{\mathrm{m}}(24)=81.7(1\pm13.2\%)$. \\
  \item The prediction of $R_{\mathrm{m}}$ based on the
relationship between $R_{\mathrm{m}}$ and $aa_{\mathrm{min}}$
depends greatly on the accurate measurement of $aa$. If the
suggested error in $aa$ (3 nT) before 1957 is corrected,   
the correlation coefficient between $R_{\mathrm{m}}$ and
$aa_{\mathrm{min}}$ ($r=0.94$) will be slightly higher, and the
peak size of cycle 24 will be predicted much lower,
$R_{\mathrm{m}}(24)=52.5\pm13.1$.
\end{enumerate}

\section*{Acknowledgments}
The author is grateful to the anonymous referee for suggestive
comments. This work is supported by National Natural Science
Foundation of China (NSFC) through grants 10973020, 40890161 and
10921303, and National Basic Research Program of China through
grant No. 2011CB811406.


\end{document}